\documentclass[aps,prl,twocolumn,superscriptaddress,tight]{revtex4-2}
\usepackage{amssymb}
\usepackage{color}
\usepackage{graphicx}
\usepackage{amsmath}
\usepackage{times}
\usepackage{bm}
\usepackage{float}
\makeatletter

\begin{document}
\title{Inferring synchronizability of networked heterogenous oscillators with machine learning}
\author{Liang Wang}
\affiliation{School of Physics and Information Technology, Shaanxi Normal University, Xi'an 710062, China}
\author{Huawei Fan}
\affiliation{School of Science, Xi'an University of Posts and Telecommunications, Xi'an 710062, China}
\author{Yafeng Wang}
\affiliation{Nonlinear Research Institute, Baoji University of Arts and Sciences, Baoji 721016, China}
\author{Jian Gao}
\affiliation{School of Science, Beijing University of Posts and Telecommunications, Beijing 100876, China}
\author{Yueheng Lan}
\affiliation{School of Science, Beijing University of Posts and Telecommunications, Beijing 100876, China}
\author{Jinghua Xiao}
\affiliation{School of Science, Beijing University of Posts and Telecommunications, Beijing 100876, China}
\author{Xingang Wang}
\email{wangxg@snnu.edu.cn}
\affiliation{School of Physics and Information Technology, Shaanxi Normal University, Xi'an 710062, China}

\begin{abstract}
In the study of network synchronization, an outstanding question of both theoretical and practical significance is how to allocate a given set of heterogenous oscillators on a complex network in order for improving the synchronization performance. Whereas methods have been proposed to address this question in literature, the methods are based on accurate models describing the system dynamics, which, however, are normally unavailable in realistic situations. Here we show that this question can be addressed by the model-free technique of feed-forward neural network (FNN) in machine learning. Specifically, we measure the synchronization performance of a number of allocation schemes and use the measured data to train a machine. It is found that the trained machine is able to not only infer the synchronization performance of {\it any} new allocation scheme, but also find from a huge amount of candidates the optimal allocation scheme for synchronization.  
\end{abstract}

\maketitle

{\it Background.--} Many real-world systems can be represented by complex networks of coupled oscillators~\cite{REV:RA}, in which an interesting phenomenon is that in some circumstances the oscillators can be self-organized into coherent states of synchronized motions~\cite{REV:JAA}. For its important implications to the functionality and performance of many realistic systems, network synchronization has been broadly interested and extensively studied by researchers from different fields in the past decades~\cite{REV:Arenas,REV:FAR,REV:SB2016}. In exploring network synchronization, one of the central questions concerns the improvement of network synchronization performance, namely synchronizability, by a slight change of the system properties~\cite{NetOptimization:2007}, e.g., introducing shortcut links or adjusting the coupling schemes~\cite{SW:1998,NetSyn:Pecora2002,Syn:Motter2005,XGW:2007,ExpSyn:2011,NetSensitive,FHW:PRE2018,SynOptimization2022}. For complex network  consisting of heterogeneous oscillators, an additional approach to improve the network synchronizability is reallocating the oscillators according to the network topology~\cite{ExpSyn:2011,SynOpt:MB,SynOpt:WY,Powergrid:Motter2013,SynOpt:ZM,SynOpt:Jfunction,SynOpt:Deng2016,SynOpt:PSS2017}. Whereas strategies have been proposed in literature on how to reallocate the oscillators optimally and the efficiency of the strategies have been justified in different systems, the existing studies rely on a prior knowledge of the network models~\cite{SynOpt:MB,SynOpt:WY,Powergrid:Motter2013,SynOpt:ZM,SynOpt:Jfunction,SynOpt:Deng2016,SynOpt:PSS2017}, including the network structure, the coupling function and the oscillator dynamics, which is normally unattainable in practice. Moreover, while evidences suggest that there might exist a one-to-one correspondence between oscillator allocation and network synchronizability, the mapping function is too complicated to be given explicitly and the searching of the optimal allocation relies on still large-scale simulations. The purpose of our present work is to introduce a model-free technique in machine learning, namely feed-forward neural network (FNN)~\cite{FNN:DS1997,Book:DL,RC:ZHScienceChina2021}, to learn from measured data the mapping function between oscillator allocation and network synchronizability, and utilize the trained machine to find the optimal allocation for synchronization.

{\it Problem description.--} We study the synchronization behaviors of coupled nonidentical phase oscillators described by the generalized Kuramoto model~\cite{REV:JAA} 
\begin{equation}
\label{model}
\dot{\theta}_{i}=\omega_{i}+\frac{\varepsilon}{d_{i}}\sum\limits_{j=1}^{N} a_{ij}\sin(\theta_{j}-\theta_{i}),
\end{equation}
with $i$, $j$ =1, ..., $N$ the oscillator (node) indices and $N$ the network size. $\theta_{i}$ and $\omega_{i}$ represent, respectively, the phase and natural frequency of the $i$th oscillator. The coupling relationship of the oscillators, i.e., the network structure, is encoded in the adjacency matrix $\mathbf{A}=\{a_{ij}\}_{N\times N}$, with $a_{ij}=a_{ji}=1$ if there is a link between nodes $i$ and $j$, and $a_{ij}=a_{ji}=0$ otherwise. $d_{i}=\sum_{j}a_{ij}$ denotes the degree of node $i$, and $\varepsilon$ is the uniform coupling strength. The network structure is generated by the Erd\"{o}s-R\'{e}nyi model with connecting probability $p$~\cite{REV:RA}. In our studies, we set $N=20$ and $p=0.5$. The natural frequencies of the oscillators are randomly chosen within the range $(-1,1)$, which, once chosen, will be fixed, but the oscillators are allowed to be reallocated freely on the network. The total number of allocation schemes is huge ($N!$), and different allocations give different synchronization performance~\cite{ExpSyn:2011,SynOpt:MB,SynOpt:WY,Powergrid:Motter2013,SynOpt:ZM,SynOpt:Jfunction,SynOpt:Deng2016,SynOpt:PSS2017}. The central question we ask is, given that the synchronization performance of a number of allocation schemes is known, can we predict the synchronization performance of a new allocation scheme without knowing the network dynamics?

\begin{figure}[t]
\begin{center}
\includegraphics[width=0.75\linewidth]{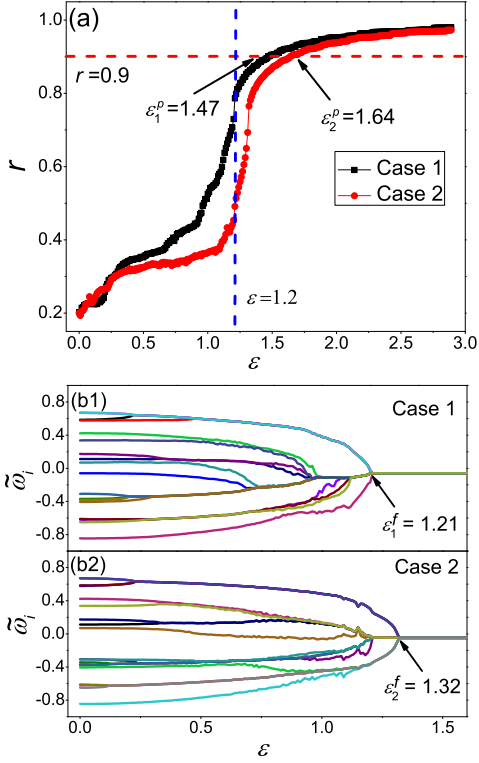}
\centering\caption{Dependence of network synchronization performance on oscillator allocation. Shown are the transitions from desynchronization to phase (a) and frequency (b) synchronization for two random oscillator allocations on a random network containing $N=20$ nodes. $r$: the synchronization order parameter; $\varepsilon^p$: the critical coupling where $r>r_c=0.9$; $\varepsilon^f$: the critical coupling where the effective frequencies of the oscillators, $\{\tilde{\omega}_i\}$, become identical.} \label{fig1}
\end{center}
\end{figure}

To show the influences of oscillator allocation on synchronization performance, we choose two random allocations and, by solving Eq.~(\ref{model}) numerically, plot in Fig.~\ref{fig1}(a) the variation of the synchronization order parameter, $r$, with respect to the coupling strength, $\varepsilon$. Here synchronization order parameter is defined as $r=|\sum_j e^{i\theta_j}|/N$, with $j=1,\ldots,N$ the oscillator index. We have $r\in (0,1)$, with larger $r$ representing stronger synchronization~\cite{REV:RA}. Figure~\ref{fig1}(a) shows that the two order parameters are close for weak couplings ($\varepsilon<0.5$), but are separated from each other for strong couplings [$\varepsilon\in (0.5,2.0)$]. Defining phase synchronization as the point where $r$ exceeds $r_c=0.9$, the critical couplings for phase synchronization are $\varepsilon_1^p\approx 1.47$ and $\varepsilon_2^p\approx 1.64$ for the first and second allocations, respectively. Clearly, in terms of phase synchronization, the first allocation outperforms the second one. Denoting $\bm{\Omega}_l$ as a specific allocation of the oscillators and $\varepsilon_l^p$ the corresponding critical coupling for phase synchronization, we have for each allocation a synchronization pair (data point) $(\bm{\Omega}_l,\varepsilon_l^p)$. Assuming that a number ($\tilde{M}\gg1$) of such synchronization pairs are available (obtained by model simulations or experiments), our {\it first objective} is to infer by machine the critical coupling for a new allocation to reach phase synchronization. 

By the same allocation schemes studied in Fig.~\ref{fig1}(a), we plot in Fig.~\ref{fig1}(b) the transitions of the network dynamics from the perspective of frequency synchronization~\cite{REV:RA}. Frequency synchronization is featured by the identical effective frequencies of the oscillators. The effective frequency of oscillator $i$ is $\tilde{\omega}_i=\langle\dot{\theta}_i\rangle_T$, with $\dot{\theta}_i$ the instant angular frequency of oscillator $i$ and $\langle\ldots\rangle_T$ the time-average function. Figure~\ref{fig1}(b) shows that for the first allocation, the effective frequencies of the oscillators become identical at the critical coupling $\varepsilon_1^f\approx 1.21$, while for the second allocation the critical coupling is $\varepsilon_2^f\approx 1.32$. Comparing to the results of phase synchronization shown in Fig.~\ref{fig1}(a), we have $\varepsilon^f<\varepsilon^p$ for both allocations. This phenomenon is understandable, as the motions of the oscillators are more strongly correlated in phase synchronization than in frequency synchronization~\cite{REV:FAR}. Assuming that a large number of frequency-based synchronization pairs of the form $(\bm{\Omega}_l,\varepsilon_l^f)$ are available, the {\it second objective} of our present work is to infer by machine the critical coupling, $\varepsilon^f$, for a new allocation scheme to achieve frequency synchronization.

Once the capability of the machine in inferring the synchronization performance of new allocation schemes has been confirmed, we shall then utilize the machine to find from a large number of candidates the optimal allocation giving the best synchronization performance, which is the {\it third objective} of our present work. Besides the critical couplings ($\varepsilon^p$ and $\varepsilon^f$), another approach to evaluate the synchronization performance of different allocations is comparing their synchronization order parameters under the same coupling strength. Taking the results in Fig.~\ref{fig1}(a) as an example, at the intermediate coupling strength $\varepsilon=1.2$, the order parameter of the first allocation is about $0.8$, while for the second allocation the order parameter is only about $0.5$. Denoting $(\bm{\Omega}_l,r_l)_{\varepsilon}$ as the measured data in this case and assuming that a number of such measurements are available, the {\it fourth objective} of our present work is to infer by machine the order parameter of a new allocation scheme under a specific coupling strength.

{\it FNN technique.--} The FNN adopted in our studies consists of three modules~\cite{FNN:DS1997,Book:DL,RC:ZHScienceChina2021}: an input layer, a series of $L$ hidden layers, and an output layer. The input layer is characterized by the matrix $\bm{W}_{in}\in\mathbb{R}^{n_1\times N}$, which couples the input vector $\bm{\Omega}_l=[\omega_1,\ldots,\omega_N]^T$ (i.e., a specific allocation of the oscillators on the network) to the $1$st hidden layer by the operation
\begin{equation}
\label{input}
\bm{x}_{1}=f(\bm{W}_{in}\bm{\Omega}_l+\bm{b}_{in}).
\end{equation}
Here, $\bm{x}_{1}\in\mathbb{R}^{n_1}$ is the state vector of the $1$st hidden layer, $n_1$ denotes the size of the $1$st hidden layer, $\bm{b}_{in}\in\mathbb{R}^{n_1}$ is the bias vector associated to the input layer, and $f(x)$ is the activation function. The signals are then propagated to the other hidden layers in sequence as
\begin{equation}
\label{hidden}
\bm{x}_{h}=f(\bm{W}_{h}\bm{x}_{h-1}+\bm{b}_{h}),
\end{equation}
where $h=2,\ldots,L$ is the index of the hidden layers, $\bm{W}_h\in\mathbb{R}^{n_h\times n_{h-1}}$ is the coupling matrix between layers $h-1$ and $h$, $n_h$ is the size of the $h$th hidden layer, and $\bm{b}_h\in\mathbb{R}^{n_h}$ is the bias vector associated to the $h$th hidden layer. After the $L$th hidden layer is activated, we generate the output
\begin{equation}
\label{output}
y=f(\bm{W}_{out}\bm{x}_L+b_{out}),
\end{equation}
with $y$ a scalar, $\bm{W}_{out}\in\mathbb{R}^{1\times n_L}$ the output matrix, and $b_{out}$ the output bias. The hyperparameters of FNN are the number of hidden layers, $L$, and the number of nodes contained in each hidden layer, $\{n_h\}$, with $h=1,\ldots,L$. The model parameters of FNN include the coupling matrices ($\bm{W}_{in}$, $\bm{W}_h$ and $\bm{W}_{out}$) and the bias vectors ($\bm{b}_{in}$, $\bm{b}_h$ and $b_{out}$), which are to be learned from the measured data through a training process. In our studies, we adopt the all-to-all coupling strategy for nodes in the neighboring layers, and use the rectified linear activation function (ReLU) to update the node states~\cite{Book:DL}.

The implementation of FNN consists of three phases: the training phase, the validation phase, and the prediction phase. In the training phase, the input data are the oscillator allocations $\{\bm{\Omega}\}_{l=1}^M$, and the mission of the training is to obtain the set of model parameters (i..e. the coupling matrices and bias vectors) such that the outputs given by Eq.~(\ref{output}) are as close as possible to the true values, which are the critical couplings ($\varepsilon^p$ or $\varepsilon^f$) for objectives $1$ and $2$ and is the synchronization order parameter ($r$) for objective $4$. In our studies, this is done by the conventional back propagation algorithm with the help of the ADAM (adaptive moment estimation) optimizer in TensorFlow~\cite{BP:1986}. After training, the performance of the machine is validated by a new dataset different from the one used in the training phase. This process of training and evaluation is repeated for a number of rounds, while in each round the structure of FNN is reconstructed according to a unique set of hyperparameters. By the strategy of Bayesian optimization~\cite{Book:DL}, we then find the set of hyperparameters giving the best performance on the validating dataset, which completes the validation phase. Finally, in the prediction phase, with the optimal hyperparameters obtained in the validation phase and the model parameters obtained in the training phase, we drive the machine with a completely new oscillator allocation, and the output of the machine gives the prediction. (See Supplementary for more details about FNN.) 

\begin{figure}[btp]
\includegraphics[width=\linewidth]{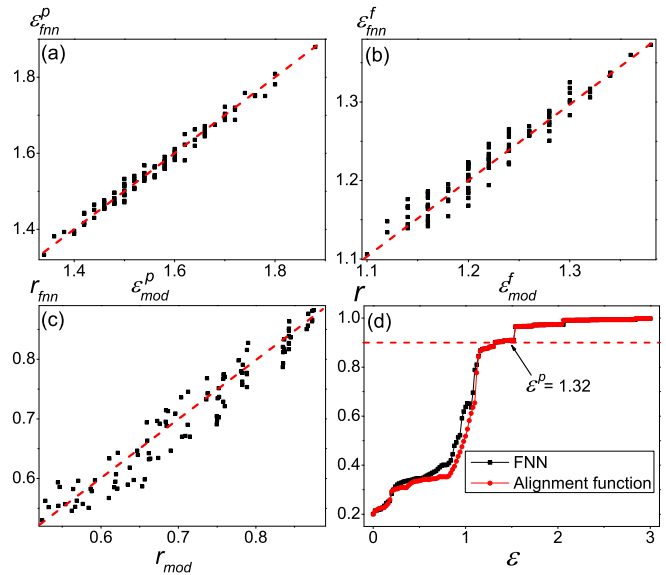}
\caption{Inferring network synchronizability with machine. (a) Inference of the critical coupling, $\varepsilon^p$, for phase synchronization. (b) Inference of the critical coupling, $\varepsilon^f$, for frequency synchronization. (c) Inference of the synchronization order parameter for the coupling strength $\varepsilon=1.2$. $\varepsilon_{fnn}$ and $\varepsilon_{mod}$ denote, respectively, the results predicted by the machine and the results obtained by model simulations. Red dashed lines are diagonal lines. (d) Synchronization performance of the optimal allocations obtained by FNN (black squares) and by the method of alignment function (red discs).}
\label{fig2}
\end{figure}

{\it Results.--} We start by showing the performance of FNN in inferring the critical coupling, $\varepsilon^p$, for phase synchronization. For simplicity, we consider here the scenario of constrained frequency allocations~\cite{SynOpt:Jfunction}. That is, only a fraction of the oscillators are reallocated on the network, while the locations of the other oscillators are fixed. The $m$ relocatable oscillators are chosen by random, and the exchange of any pair of them generates a new allocation. In our studies, we set $m=10$ and generate in total $\tilde{M}=800$ random allocations. For each allocation, we first obtain by simulations the variation of the order parameter, $r$, with respect to the coupling strength, $\varepsilon$, by the increment $\Delta\varepsilon=0.02$, and then find the critical coupling, $\varepsilon^p$, where $r$ exceeds $r_c=0.9$. We obtain in total $\tilde{M}=800$ data points of the form $(\bm{\Omega}_l,\varepsilon^p_l)$, among which $M=500$ points are used in the training phase for obtaining the model parameters and $M'=300$ points are used in the validation phase for optimizing the machine hyperparameters (see Supplementary for the statistical properties of the dataset and the influence of the data properties on machine performance). In this application, the optimal hyperparameters are $L=1$ and $n_1=100$. That is, the FNN contains only one hidden layer and the size of the hidden layer is $100$. Shown in Fig.~\ref{fig2}(a) are the critical couplings predicted by the machine for $100$ new allocations that are not included in the training and validating dataset. We see that the results predicted by the machine are in good agreement with the results of model simulations. To evaluate the overall predicting performance of the machine, we introduce the normalized error $e^p=|1-\varepsilon_{fnn}^p/\varepsilon_{mod}^p|$ for each prediction, with $\varepsilon_{fnn}^p$ and $\varepsilon_{mod}^p$ denoting, respectively, the critical couplings predicted by the machine and obtained by model simulation, and calculate the averaged predicting error, $\langle e^p\rangle$, for $K=1000$ new random allocations. We have $\langle e^p\rangle\approx 7\times 10^{-3}$. Indeed, the machine is able to infer the critical coupling for phase synchronization with a high precision. 

We next employ the FNN technique to predict the critical coupling for frequency synchronization. In doing this, we first regenerate the dataset by model simulations, and then use the measured data to construct and train a new machine. The data point has the form $(\bm{\Omega}_l, \varepsilon^f_l)$, with $\bm{\Omega}_l$ the $l$th allocation scheme and $\varepsilon^f_l$ the corresponding critical coupling. Again, we generate in total $\tilde{M}=800$ data points ($M=500$ points for training the model parameters and $M'=300$ points for finding the optimal hyperparameters). In this application, the optimal hyperparameters are $L=1$ and $n_1=100$. We consider still the scenario of constrained allocations, with the number of relocatable oscillators being $m=10$. Figure~\ref{fig2}(b) shows the prediction performance of the new machine for $100$ random allocations. We see that the critical couplings are accurately predicted by the machine in all cases. In this application, the predicting error averaged over $K=1000$ random allocations is about $\langle e^f\rangle=1\times 10^{-2}$, justifying hence the capability of the machine in inferring frequency synchronization. 

\begin{table*}[tbp]
\caption{\label{tab1} The performance of FNN in other network models. $N$: the network size. $m$: the number of relocatable oscillators. $\tilde{M}$: the number of data points obtained by model simulations in inferring phase synchronization. $M$: the number of data points used in machine training. $\langle e^p\rangle$: averaged prediction error for phase synchronization. $\langle e^f\rangle$: averaged prediction error for frequency synchronization. $\langle e\rangle_r$: averaged prediction error for synchronization order parameter under the coupling strength $\varepsilon_r$. Errors are averaged over $K=1000$ random allocations.}
\begin{ruledtabular}
\begin{tabular}{ccccccc}
\mbox{Models} & \mbox{Nodal dynamics} & $N/m$ & $\tilde{M}/M$ & $\langle e^p\rangle$ & $\langle e^f\rangle$ & $(\langle e\rangle_r, \varepsilon_r)$ \\
\hline
Scale-free network~\cite{BA:1999} & Phase oscillator & $100/10$ & $1000/700$ & $3.5\times 10^{-3}$ & $4\times 10^{-3}$ & $(6\times 10^{-2},1.38)$ \\
Nepal power grid~\cite{Pecora2014} & Phase oscillator & $15/15$ & $1500/1000$ & $2.3\times 10^{-2}$ & $3\times 10^{-2}$ & $(5\times 10^{-2},1.0)$ \\
Cat brain network~\cite{CatBrain} & Phase oscillator & $53/10$ & $1200/800$ & $5\times 10^{-3}$ & $6\times 10^{-3}$ & $(1.6\times 10^{-2},1.3)$ \\
Nematode neuronal network~\cite{Elegans} & Phase oscillator & $297/10$ & $2000/1500$ & $3.2\times 10^{-3}$ & $2.7\times 10^{-3}$ & $(1.7\times 10^{-2},1.48)$ \\
Complex random network & R\"{o}ssler oscillator~\cite{Rossler} & $20/10$ & $2000/1500$ & $1.4\times 10^{-2}$ & $2.4\times 10^{-2}$ & $(2.7\times 10^{-2},0.025)$ \\
Complex ecological network~\cite{FoodNet} & Foodweb oscillator~\cite{Model:foodweb} & $82/10$ & $2500/2000$ & $2.8\times 10^{-2}$ & $2.9\times 10^{-2}$ & $(1.2\times 10^{-2},0.5)$ \\
\end{tabular}
\end{ruledtabular}
\end{table*}

Having shown the capability of the FNN technique in predicting phase and frequency synchronization, we continue to employ the technique as an efficient tool to find the optimal allocation for synchronization among a huge amount of candidates (objective 3 in our present work). To study, we first generate $10000$ allocations by exchanging the ($m=10$) relocatable oscillators randomly, and then use the machine trained for predicting phase synchronization to estimate the critical coupling $\varepsilon^p$ for each allocation. The optimal allocation is identified as the one with the smallest $\varepsilon^p$, which is about $1.32$ among the $10000$ candidates. As the machine has been trained already and predictions are made by a mapping function, the searching of the optimal allocation is accomplished in seconds. By numerical simulations, we plot in Fig.~\ref{fig2}(d) the variation of $r$ with respect to $\varepsilon$ for the optimal allocation identified by the machine. It is seen that $r$ exceeds $r_c=0.9$ at $\varepsilon^p\approx1.32$, which is exactly the critical coupling predicted by the machine. As a benchmark to evaluate the synchronization performance of the optimal allocation, we compare it to the optimal allocation obtained by the method of synchrony alignment function proposed in Ref.~\cite{SynOpt:Jfunction}. The alignment function is calculated as $J(\bm{\Omega},\bm{L})=(1/N)\sum_{i=2}^{N}\lambda^{-2}_{i}\langle\bm{\nu}_{i},\bm{\Omega}\rangle^{2}$, with $\bm{L}=\bm{D}-\bm{A}$ the Laplacian matrix, $\bm{D}$ the diagonal matrix encoding the node degrees, $\{\lambda\}_{i=2}^N$ the nonzero eigenvalues of $\bm{L}$, and $\{\bm{\nu}_{i}\}_{i=2}^N$ the corresponding eigenvectors. We apply the function to the same set of candidates, and the one with the smallest value of $J$ is identified as the optimal allocation. The synchronization performance of the new optimal allocation is also shown in Fig.~\ref{fig2}(d). We see that the order parameters of the two optimal allocations are close for weak couplings ($\varepsilon<1.2$) and are identical for strong couplings ($\varepsilon>1.2$). Additional analysis has been conducted to check the properties of the optimal allocation identified by the machine, which shows that, similar to the results of alignment function, the degrees and frequencies are positively correlated and the frequencies of the neighboring nodes are negatively correlated (see Supplementary for details).  

We finally employ the FNN technique to infer the synchronization order parameter (objective 4). Fixing $\varepsilon=1.2$, we obtain by simulations the order parameter for $\tilde{M}=1000$ random allocations and, using $(\bm{\Omega}_l,r_l)_{l=1}^{\tilde{M}}$ as the inputs, optimize and train a new machine. The hyperparameters of this new machine are $L=3$ and $(n_1,n_2,n_3)=(150,200,50)$. The performance of the new machine in predicting the synchronization order parameter of $100$ new random allocations is shown in Fig.~\ref{fig2}(c). We see that the order parameters predicted by the machine agree with the true results very well. The normalized prediction error averaged over $K=1000$ random allocations is $\langle e\rangle_r\approx 3.7\times 10^{-2}$, which is slightly larger than the prediction errors of the critical couplings ($7\times 10^{-3}$ for $\varepsilon^p$ and $1\times 10^{-2}$ for $\varepsilon^f$). This arouses our interest about the dependence of prediction performance on network synchronization degree. Intuitively, the larger the order parameter, the stronger the correlation between the oscillators, the easier the machine to learn from data the mapping function between oscillator allocation and network synchronizability, and the more accurate the inference. To verify this conjecture, we analyze again the simulating results obtained in inferring phase synchronization (i.e., the synchronization transitions of $\tilde{M}=800$ random allocations) and find the critical couplings where $r$ exceeds $r_c=0.6$. With the newly obtained dataset, we construct and train a new machine (the hyperparameters are $L=2$ and $n_1=n_2=100$), and then use the trained machine to predict the order parameters for $K=1000$ new random allocations. The averaged prediction error is $\langle e\rangle_r\approx 3\times 10^{-2}$. To verify the conjecture further, we set $r_c=0.3$ and check again the prediction performance (the hyperparameters are $L=1$ and $n_1=100$). In this case, we have $\langle e\rangle_r\approx 0.2$. Indeed, the prediction performance is improved by increasing the synchronization degree.  

{\it Discussions.--} The capability and performance of the FNN technique have been verified in other network models, including networks of different sizes and topological features, networks consisting of heterogeneous chaotic oscillators, and also the scenario of unconstrained allocations. The results are summarized in Tab.~\ref{tab1} (details are given in Supplementary). We see that, while FNN performs well in all the models, its performance is dependent on the complexity of the system dynamics. Generally, the larger the network size, the larger the FNN; the lower the synchronization degree, the worse the predictions; and, comparing to phase oscillators, the inference of chaotic oscillators is more difficult. Besides inferring the synchronization performance of relocated oscillators, the trained machine is also capable of inferring the synchronization performance of a new set of oscillators with different natural frequencies, showing preliminary the feature of transfer learning (please see Supplementary for the details). We finally note that the FNN technique is distinguished from the existing methods by its model-free feature, but not the computational efficiency, as the acquisition of the training data relies on still simulations. In addition, while evidences indicate that the machine has learned successfully from data the mapping function between oscillator allocation and synchronization performance, the function is not explicitly given, i.e., the machine is working as a ``black box". 

\begin{acknowledgments}
This work was supported by the Natural Science Foundation of China under Grant Nos.~11875182 and~12275165. XGW was also supported by the Fundamental Research Funds for the Central Universities under Grant No.~GK202202003.

\end{acknowledgments}

\end{document}